\documentclass[english,prd,twocolumn,superscriptaddress,nofootinbib,preprintnumbers]{revtex4}
\usepackage[latin1]{inputenc}
\usepackage{graphicx}
\usepackage{amssymb}

\usepackage{amsfonts}
\usepackage{dcolumn}
\usepackage{hyperref}

\def\be{\begin{equation}}
\def\ee{\end{equation}}
\def\ba{\begin{eqnarray}}
\def\ea{\end{eqnarray}}
\def\bs{\begin{subequations}}
\def\es{\end{subequations}}

\newcommand{\rd}{{\rm d}}

\usepackage{color}

\makeatother

\usepackage{babel}
\makeatother
\begin{document}

\title{The dispersion of growth of matter perturbations in $f(R)$ gravity}

\author{Shinji Tsujikawa}
\affiliation{Department of Physics, Faculty of Science, Tokyo University of Science, 
1-3, Kagurazaka, Shinjuku-ku, Tokyo 162-8601, Japan}

\author{Radouane Gannouji}
\affiliation{IUCAA, Post Bag 4, Ganeshkhind, Pune 411 007, India}

\author{Bruno Moraes}
\affiliation{Lab. de Physique Theorique et Astroparticules, CNRS
Universite Montpellier II, France}

\author{David Polarski}
\affiliation{Lab. de Physique Theorique et Astroparticules, CNRS
Universite Montpellier II, France}

%
\begin{abstract}

We study the growth of matter density perturbations $\delta_m$ 
for a number of viable $f(R)$ gravity models that satisfy 
both cosmological and local gravity constraints, 
where the Lagrangian density $f$ is a function of the Ricci scalar $R$.
If the parameter $m \equiv Rf_{,RR}/f_{,R}$ today is larger than 
the order of $10^{-6}$, linear perturbations relevant to the 
matter power spectrum evolve with a growth rate 
$s\equiv \rd \ln \delta_m/\rd \ln a$ ($a$ is the scale factor)
that is larger than in the $\Lambda$CDM model.
We find the window in the free parameter space of our models 
for which spatial dispersion of the growth index 
$\gamma_0 \equiv \gamma(z=0)$ 
($z$ is the redshift) appears in the range of values 
$0.40 \lesssim \gamma_0 \lesssim 0.55$, 
as well as the region in parameter space for which there is 
essentially no dispersion and $\gamma_0$ converges to values 
around $0.40 \lesssim \gamma_0 \lesssim 0.43$. These latter 
values are much lower than in the $\Lambda$CDM model. 
We show that these unusual dispersed or converged spectra
are present in most of the viable $f(R)$ models with $m(z=0)$
larger than the order of $10^{-6}$. These properties will be essential in 
the quest for $f(R)$ modified gravity models using future 
high-precision observations and they confirm the possibility 
to distinguish clearly most of these models from 
the $\Lambda$CDM model.

\end{abstract}

\date{\today}

\maketitle


\section{Introduction}

The origin of dark energy (DE) responsible for the cosmic acceleration
today has been a lasting mystery \cite{review}.
Although a host of independent observational data have supported the 
existence of DE over the past ten years, no strong evidence was found 
yet implying that dynamical DE models are better than a  
cosmological constant $\Lambda$. 
A first step towards understanding the origin of DE would be
to detect some clear deviation from the $\Lambda$CDM model 
observationally and experimentally.

Models such as quintessence \cite{quin} based on minimally coupled
scalar fields provide a dynamical equation of state 
of DE different from $w_{\rm DE}=-1$. 
Still it is difficult to distinguish these models
from the $\Lambda$CDM model in current observations
pertaining to the cosmic expansion history only
(such as the supernovae Ia observations).
Even if we consider the evolution of matter perturbations $\delta_m$ 
in these models, the growth rate of $\delta_m$ is similar to 
that in the $\Lambda$CDM model.
Hence one cannot generally expect large differences with the
$\Lambda$CDM model at both the background and the perturbation levels.

There is another class of DE models in which gravity is modified 
with respect to General Relativity (GR). The simplest one would be 
the so-called $f(R)$ gravity where the Lagrangian density $f$ is a 
function of the Ricci scalar $R$ \cite{fRearly}.
The basic idea is that gravity is modified on cosmological 
scales when $R$ is of the order of $H_0^2$ ($H_0$ is the Hubble parameter
today), while Newtonian gravity is recovered in the region of high 
density ($R \gg H_0^2$). A number of viable $f(R)$ models have been 
constructed in this spirit \cite{AGPT,Li,AT08,Hu,Star,Appleby,Tsuji,N007,Linder}. 
Since the law of gravity is modified in $f(R)$ models, 
we can in principle expect large differences with the $\Lambda$CDM model 
in cosmological observations \cite{obserpapers,lensing} and 
in laboratory tests \cite{lgcpapers,Navarro,CT} compared to quintessence models. 

From a cosmological point of view viable $f(R)$ models are similar to 
the $\Lambda$CDM model during the radiation and deep matter eras (``GR regime''), 
but important observable deviations from the $\Lambda$CDM model appear 
at late times as the model evolves towards what we call here the 
``scalar-tensor regime'' [see below after Eq.~(\ref{mapprox})].
A useful quantity that characterizes this deviation is 
$m=Rf_{,RR}/f_{,R}$ \cite{AGPT}, 
where $f_{,R} \equiv \partial f/\partial R$ and 
$f_{,RR} \equiv \partial^2 f/\partial R^2$.
In order to satisfy local gravity constraints we require that $m$ is 
much smaller than 1 for $R \gg H_0^2$, e.g., 
$m(R) \lesssim 10^{-15}$ for $R \approx 10^5 H_0^2$ \cite{AT08,CT,Tavakol}.
Meanwhile, in order to see appreciable deviation from the $\Lambda$CDM 
model at the background level of cosmological evolution, 
the parameter $m$ needs to grow to the of order at least 0.01-0.1 today. 
The models proposed in Refs.~\cite{Hu,Star,Appleby,Tsuji,Linder} 
are constructed to realize this fast transition of $m$. Actually as we will 
see, the quantity $m$ is related to the (critical) scale 
$\sim M^{-1} = \left( 3m/R \right)^{1/2}$ below which 
modifications of gravity are felt by the matter perturbations. 
For increasing $m$ and for decreasing $R$, 
this critical scale gets larger. 

The modified evolution of the matter density perturbations $\delta_m$ 
provides an important tool to distinguish $f(R)$ models, and generally 
modified gravity DE models, from DE models inside GR and in particular 
from the $\Lambda$CDM model \cite{obserpapers}.
In fact the effective gravitational ``constant'' $G_{\rm eff}$ which appears 
in the source term driving the evolution of matter perturbations can change 
significantly relative to the gravitational constant $G$ in the usual GR 
regime, i.e. $G_{\rm eff} \simeq (4/3)G$ (the ``scalar-tensor'' regime). 
Then the evolution of perturbations during the 
matter era changes from $\delta_m \propto t^{2/3}$ to 
$\delta_m \propto t^{(\sqrt{33}-1)/6}$, where 
$t$ is the cosmic time \cite{Star,Tsuji}.

A useful way to describe the perturbations is to write the growth 
function $s=\rd \ln \delta_m/\rd \ln a$ 
as $s=(\Omega_m)^{\gamma}$, where $\Omega_m$ is the 
density parameter of non-relativistic matter.
As well-known one has $\gamma_0\equiv \gamma(z=0)\simeq 0.55$ 
\cite{Wang,Lindergamma} in the $\Lambda$CDM model.
It was emphasized that while $\gamma$ is quasi-constant in standard 
(non-interacting) DE models inside GR with $\gamma_0 \simeq 0.55$, this 
needs not be the case in modified gravity models, in particular large 
slopes can appear \cite{PG08} 
(see Refs.~\cite{gammapapers} for more related works). 
For the model proposed by Starobinsky \cite{Star}
it was found in Ref.~\cite{GMP09} that the present value of 
the growth index $\gamma_0$ can be as small as 
$\gamma_0 =$0.40-0.43 
while large slopes are obtained.
This allows to clearly discriminate this model from $\Lambda$CDM.  
An additional important point is whether $\gamma_0$ can exhibit some dispersion 
(scale dependence) for viable $f(R)$ DE models.  

The redshift at which the transition of perturbations occurs 
depends on the comoving wavenumber $k$.
It is then expected that the resulting matter power spectrum 
has a scale dependence for viable $f(R)$ models.
In this paper we shall study the dependence of the growth 
index $\gamma$ on scales relevant to the linear regime
of the matter power spectrum.
We consider most of viable $f(R)$ models proposed in literature
to understand general properties of the dispersion of perturbations.
This analysis will be important to distinguish between the $f(R)$ models 
and the $\Lambda$CDM model in future 
observations of galaxy clustering and weak lensing.


\section{Cosmology in $f(R)$ gravity}

We start with the action
\begin{equation}
\label{action}
S= \frac{1}{2\kappa^{2}} \int{\rm d}^{4}x\sqrt{-g}f(R)
+S_m(g_{\mu \nu}, \Psi_m)\,,
\end{equation}
where $\kappa^{2}=8\pi G$ ($G$ is bare gravitational constant),
and $S_m$ is a matter action that depends on the metric 
$g_{\mu \nu}$ and matter fields $\Psi_m$.
Since we are interested in the cosmological evolution at the late
epoch, we only consider a perfect fluid of non-relativistic matter
with an energy density $\rho_m$.
In the following we use the unit $\kappa^2=1$, but we restore the 
gravitational constant $G$ when the discussion 
becomes transparent.

In the flat Friedmann-Lema\^{i}tre-Robertson-Walker (FLRW) spacetime 
with scale factor $a(t)$, the variation of the action (\ref{action}) 
leads to the following equations
\begin{eqnarray}
3FH^{2} & = & \rho_m
+(FR-f)/2-3H\dot{F},\label{E1}\\
-2F\dot{H} & = & \rho_m
+\ddot{F}-H\dot{F}, \label{E2}
\end{eqnarray}
where $F \equiv \partial f/\partial R$, $H \equiv \dot{a}/a$, and
a dot represents a derivative with respect to the cosmic time $t$.
The Ricci scalar $R$ is expressed by the Hubble parameter $H$
as $R=6(2H^2+\dot{H})$.
In order to study the cosmological dynamics in $f(R)$ gravity, 
it is convenient to introduce the following variables
\begin{eqnarray}
x_{1}=-\frac{\dot{F}}{HF}\,,\quad 
x_{2}=-\frac{f}{6FH^{2}}\,, \quad
x_{3}=\frac{R}{6H^{2}}\,,
\label{didef}
\end{eqnarray}
together with the matter density parameter 
\begin{equation}
\tilde{\Omega}_m \equiv \frac{\rho_m}{3FH^2}=1-x_1-x_2-x_3\,.
\label{didef2}
\end{equation}
We then obtain the following dynamical equations \cite{AGPT}
\begin{eqnarray}
x_{1}' & = & -1-x_{3}-3x_{2}+x_{1}^{2}-x_{1}x_{3}\,,\label{N1}\\
x_{2}' & = & \frac{x_{1}x_{3}}{m(r)}-x_{2}(2x_{3}-4-x_{1})\,,\label{N2}\\
x_{3}' & = & -\frac{x_{1}x_{3}}{m(r)}-2x_{3}(x_{3}-2)\,,\label{N3}
\end{eqnarray}
where a prime represents a derivative with respect to $N=\ln a$, and 
\begin{equation}
m(r) \equiv \frac{Rf_{,RR}}{f_{,R}}\,,\quad
r \equiv -\frac{Rf_{,R}}{f}=\frac{x_3}{x_2}\,.
\label{mrdef}
\end{equation}

One has $m=0$ for the $\Lambda$CDM 
model ($f(R)=R-2\Lambda$), so that the quantity $m$
characterizes the deviation from the $\Lambda$CDM model.
Since $m$ is a function of $r=x_3/x_2$, the above dynamical 
equations are closed. For given functional forms of $f(R)$, the background 
cosmological dynamics is known by solving Eqs.~(\ref{N1})-(\ref{N3})
with Eq.~(\ref{mrdef}).
The matter point $P_m$ and the de Sitter point $P_{\rm dS}$
correspond to \cite{AGPT}
\begin{itemize}
\item $P_m$: $(x_1,x_2,x_3)=\left( \frac{3m}{1+m}, -\frac{1+4m}{2(1+m)^2},
\frac{1+4m}{2(1+m)} \right)$,\\
~~~~~~~$\tilde{\Omega}_m=1-\frac{m(7+10m)}{2(1+m)^2}$,~~~
$w_{\rm eff}=-\frac{m}{1+m}$\,.
\item $P_{\rm dS}$: $(x_1,x_2,x_3)=(0,-1,2)$,\\
~~~~~~~$\tilde{\Omega}_m=0$,~~~
$w_{\rm eff}=-1$\,.
\end{itemize}
We require that $m \approx 0$ to realize the matter era with 
$\tilde{\Omega}_m \approx 1$ and $w_{\rm eff} \approx 0$.
{}From the definition of $r$ in Eq.~(\ref{mrdef}) we have 
$r=-m-1$ for $P_m$, so that the matter point
corresponds to $(r, m)\approx (-1, 0)$ in the $(r, m)$ plane.
Note that the radiation point also exists around 
$(r, m)\approx (-1, 0)$ \cite{AGPT}.
The de Sitter point $P_{\rm dS}$ correspond to the line $r=-2$.

Let us next consider linear perturbations about the flat FLRW background.
In the so-called comoving gauge \cite{Hwang} where the velocity 
perturbation of non-relativistic matter vanishes, the matter perturbation 
$\delta_m=\delta \rho_m/\rho_m$ and the perturbation $\delta F$
obey the following equations in the Fourier space \cite{Hwang,Tsuji}
\begin{eqnarray}
& & \ddot{\delta}_m+\left(2H+\frac{\dot{F}}{2F}
\right) \dot{\delta}_m-\frac{\rho_m}{2F}\delta_m \nonumber \\
& &=\frac{1}{2F} \left[ \left(-6H^2+\frac{k^2}{a^2}
\right)\delta F+3H \dot{\delta F}+3 \ddot{\delta F}
\right], \label{delm} \\
& & \ddot{\delta F}+3H\dot{\delta F}+\left(\frac{k^2}
{a^2}+\frac{f_{,R}}{3f_{,RR}}-\frac{R}{3} \right)\delta F \nonumber \\
& &=\frac13 \rho_m \delta_m +\dot{F} \dot{\delta}_m\,,
\label{delF}
\end{eqnarray}
where $k$ is a comoving wavenumber.
We also define 
\begin{equation}
M^2 \equiv \frac{f_{,R}}{3f_{,RR}}=\frac{R}{3m}\,,
\label{mass}
\end{equation}
which corresponds to the mass squared of the scalar-field 
degree of freedom, the scalaron introduced in \cite{S80}, 
in the region $M^2 \gg R$. As we will see later, the quantity 
$m$ remains smaller than the order of $0.1$ in the cosmic 
expansion history and the condition $M^2 \gg R$ is largely 
satisfied at high redshifts.  

For cosmologically viable models, the variation of $F$
is small ($|\dot{F}| \ll HF$) so that the terms including $\dot{F}$
can be neglected in Eqs.~(\ref{delm}) and (\ref{delF}). 
If we neglect the oscillating mode of $\delta F$ relative to 
the mode induced by matter perturbations $\delta_m$, 
it follows that $\delta F \simeq \rho_m \delta_m/[3(k^2/a^2+M^2)]$ from 
Eq.~(\ref{delF}). For the modes deep inside the Hubble radius ($k^2/a^2 \gg H^2$) 
the dominant term in the square bracket of Eq.~(\ref{delm}) is $(k^2/a^2)\delta F$.
We then obtain the following approximate equation for matter 
perturbations \cite{Tsuji07,Tavakol}
\begin{equation}
\ddot{\delta}_m+2H\dot{\delta}_m
-4\pi G_{\rm eff} \rho_m \delta_m \simeq 0\,,
\label{mapprox}
\end{equation}
where 
\ba
G_{\rm eff} &\equiv& \frac{G}{F} \frac{4k^2/a^2+3M^2}{3(k^2/a^2+M^2)}\\
& = & \frac{G}{F}~\left[ 1 + \frac{1}{3}
         \frac{k^2/(a^2M^2)}{1 + k^2/(a^2 M^2)} \right]\,.
\label{Geff} 
\ea
Here we have restored bare gravitational constant $G$.
An equation of a similar type was found for scalar-tensor DE models 
\cite{BEPS00} with an essentially massless dilaton field. The quantity 
$G_{\rm eff}$ encodes the modification of gravity in the weak-field regime 
due to the presence of the dilaton in the case of scalar-tensor DE models, 
and of the scalaron in our $f(R)$ models. An essential difference is that 
the scalaron mass $M$ in the region of high density can be large, inducing a finite 
range $\sim M^{-1}$ for the (Yukawa type) ``fifth-force''. 

Under the linear expansion of the Ricci scalar $R$ in the weak gravity 
background of a spherically symmetric spacetime, the effective
Newtonian gravitational constant is given by \cite{lgcpapers,Navarro}
\begin{equation}
G_{\rm eff}^{(\rm N)}=\frac{G}{F}\left( 1 + \frac{1}{3}~e^{-Mr}\right)\,,
\label{GeffN}
\end{equation}
where $r$ is the distance from the center of symmetry. 

Poisson's equation in Fourier space with $G$ replaced by $G_{\rm eff}^{(\rm N)}$ 
as given in Eq.~(\ref{GeffN}) corresponds to the gravitational potential in 
real space (between two unit masses) 
$V(r) = -G_{\rm eff}^{(\rm N)}/r$.
It corresponds to the modification of gravity in the weak-field regime 
that is felt by the cosmological perturbations. 

For the validity of the linear expansion of $R$ the mass $M$ needs
to be light such that $Mr_c \ll 1$, where $r_c$ is the radius of 
a spherically symmetric body.
For small scales satisfying $r\ll M^{-1}$, the fifth-force reaches its maximal 
value and the modification of gravity reduces to a shift of the gravitational 
constant $G \to 4G/(3F)$. 
This is the regime we have called here the ``scalar-tensor'' regime. 
Cosmologically the corresponding shift in the scalar-tensor DE models 
mentioned above depends on time and occurs on all scales.  
Note that in the massive regime $Mr_c \gg 1$ the result (\ref{GeffN})
is no longer valid. It is exactly in this regime where the chameleon 
mechanism \cite{KW} begins to be at work so that the effective 
gravitational constant becomes very close to $G$ to satisfy local 
gravity constraints \cite{Navarro,Hu,AT08} (as we will see in the next section).

In the region $M^2 \gg k^2/a^2$ the cosmological effective gravitational 
constant (\ref{Geff}) reduces to the form $G_{\rm eff} \simeq G/F$.
Then the evolution of $\delta_m$ during the matter era
is given by $\delta_m \propto t^{2/3}$ (note that $F \simeq 1$
because the deviation parameter $m$ is much smaller than 1).
In the region $M^2 \ll k^2/a^2$ we have $G_{\rm eff} \simeq 4G/(3F)$, 
so that the matter perturbation evolves as 
$\delta_m \propto t^{(\sqrt{33}-1)/6}$ 
during the matter era \cite{Star}. 

The perturbation equation (\ref{mapprox}) has been derived for sub-horizon modes
under the neglect of the oscillating mode (such as the term $\ddot{\delta F}$).
For cosmologically viable models, the solutions obtained by solving 
Eq.~(\ref{mapprox}) agrees well with full numerical 
solutions \cite{Tavakol,Cruz,Motohashi}.
In our numerical simulations, however, we shall solve the full perturbation 
equations (\ref{delm}) and (\ref{delF}) together with the background 
equations (\ref{N1})-(\ref{N3}), without relying on the approximate 
equation (\ref{mapprox}). For the numerical integration 
it is convenient to rewrite Eqs.~(\ref{delm}) and (\ref{delF}) 
in the following forms \cite{Tsuji}
\begin{eqnarray}
\label{per1nu}
& &\delta_m''+\left(x_3-\frac12 x_1 \right)\delta_m'-
\frac32 (1-x_1-x_2-x_3) \delta_m \nonumber \\
& &=\frac12 \biggl[ \left\{ \frac{k^2}{x_4^2}-6
+3x_1^2-3x_1'-3x_1 (x_3-1) \right\}
\delta \tilde{F}\nonumber \\ 
& &~~~~~~~+3(-2x_1+x_3-1)\delta \tilde{F}'+
3\delta \tilde{F}'' \biggr]\,,\\
\label{per2nu}
& & \delta \tilde{F}''+(1-2x_1+x_3)\delta \tilde{F}' 
\nonumber \\
& & +\left[ \frac{k^2}{x_4^2}-2x_3+\frac{2x_3}{m}
-x_1 (x_3+1)-x_1'+x_1^2 \right]\delta \tilde{F} 
\nonumber \\
& & =(1-x_1-x_2-x_3)\delta _m-x_1\delta_m'\,,
\end{eqnarray}
where $\delta \tilde{F} \equiv \delta F/F$, and 
the new variable $x_4 \equiv aH$ satisfies
\begin{eqnarray}
\label{x4eq}
x_4'=(x_3-1)x_4\,.
\end{eqnarray}

The growth index $\gamma$ of matter perturbations 
is defined as 
\begin{equation}
s \equiv (\Omega_m)^{\gamma}\,,
\label{gamdef}
\end{equation}
where $s\equiv \rd \ln \delta_m/\rd \ln a=\delta_m'/\delta_m$ 
and $\Omega_m$ is given by 
\begin{equation}
\Omega_m \equiv \frac{\rho_m}{3H^2}
=F \tilde{\Omega}_m\,.
\end{equation}
This choice of $\Omega_m$ comes from rewriting Eq.~(\ref{E1})
in the form $3H^2=\rho_m+\rho_{\rm DE}$, where 
$\rho_{\rm DE} \equiv (FR-f)/2-3H\dot{F}+3H^2(1-F)$ \cite{Star,GMP09}.
For viable $f(R)$ models the quantity $F$ approaches 1 in the 
asymptotic past because they are similar to the 
$\Lambda$CDM model (as we will see in the next section).
Defining $\rho_{\rm DE}$ as well as $\Omega_m$ in the above way, 
the Friedmann equations recast in their usual General Relativistic 
form for dust-like matter and DE. 


\section{Viable $f(R)$ models}
\label{viablemodel}

Let us briefly review viable $f(R)$ models that can satisfy both cosmological and 
local gravity constraints. We focus on the models in which cosmological
solutions have a late-time de Sitter attractor at $R=R_1~(>0)$.
For the viability of $f(R)$ models the following 
conditions need to be satisfied.
\begin{itemize}
\item (i) $f_{,R}>0$ for $R \ge R_1~(>0)$. This is required to avoid anti-gravity.
\item (ii) $f_{,RR}>0$ for $R \ge R_1$. This is required for the stability 
of cosmological perturbations \cite{obserpapers}, 
for the presence of a matter era \cite{APT}, 
and for consistency with local gravity tests \cite{lgcpapers}.
\item (ii) $f(R) \to R-2\Lambda$ for $R \gg R_0$, where 
$R_0$ is the Ricci scalar today.
This is required for consistency with local gravity tests \cite{lgcpapers}
and for the presence of radiation and matter eras \cite{AGPT}.
\item (iv) $0<m(r=-2)<1$. This is required for the stability of
the late-time de Sitter point \cite{Sch,Fara,AGPT}.
\end{itemize}
The conditions (i) and (ii) mean that $m=Rf_{,RR}/f_{,R}>0$ for $R \ge R_1$.
The trajectories starting from $(r,m) \approx (-1,+0)$
to the de Sitter point on the line $r=-2$, $0<m<1$ are acceptable.
In other words, the deviation parameter $m$ is initially small ($0<m \ll 1$) 
so that the model is close to the $\Lambda$CDM model during the radiation and 
deep matter eras. The deviation from the $\Lambda$CDM model can be important 
at the late epoch. Depending on the models of $f(R)$ gravity, the parameter $m$
can grow as large as to the order of 0.1 today.

We also require the condition $0<m \ll 1$ in the region $R \gg R_0$
for consistency with local gravity constraints.
In this case the mass squared $M^2$ in Eq.~(\ref{mass}) becomes large
in the region of high density to avoid the propagation of the fifth force.
It is then possible for $f(R)$ models to satisfy local gravity constraints
under the chameleon mechanism \cite{KW}.
We introduce a new metric variable $\tilde{g}_{\mu \nu}$ and 
a scalar field $\phi$, as $\tilde{g}_{\mu \nu}=\psi g_{\mu \nu}$
and $\phi=\sqrt{3/2}\,\ln \psi$, where $\psi=F(R)$.
The action in the Einstein frame is then given by \cite{Maeda}
\begin{eqnarray}
S &=& \int {\rm d}^4 x \sqrt{-\tilde{g}} \left[ \tilde{R}/2
-(\tilde{\nabla} \phi)^2/2-V(\phi) \right] \nonumber \\
& &+S_m (\tilde{g}_{\mu \nu}e^{2Q \phi}, \Psi_m)\,,
\end{eqnarray}
where 
\begin{equation}
Q=-\frac{1}{\sqrt{6}}\,,\quad
V=\frac{R(\psi)\psi-f}{2\psi^2}\,.
\end{equation}
The scalar field degree of freedom $\phi$ has a constant 
coupling $Q$ with non-relativistic matter in the Einstein frame.

In a spherically symmetric spacetime of the Minkowski background
the field $\phi$ obeys the following equation in the Einstein frame
\begin{equation}
\frac{{\rm d}^2\phi}{{\rm d} \tilde{r}^2}+\frac{2}{\tilde{r}}
\frac{{\rm d}\phi}{{\rm d}\tilde{r}}=\frac{{\rm d} V_{\rm eff}}
{{\rm d}\phi}\,,
\end{equation}
where $\tilde{r}$ is the distance from the center of symmetry and 
\begin{equation}
V_{\rm eff}(\phi)=V(\phi)+e^{Q\phi} \rho^*\,.
\end{equation}
Here $\rho^*=e^{3Q \phi}\rho_m$ is a conserved matter density 
in the Einstein frame \cite{KW}.
For $f(R)$ models ($Q=-1/\sqrt{6}$) the effective potential 
$V_{\rm eff}(\phi)$ possesses a minimum for $V_{,\phi}(\phi)>0$.
For a spherically symmetric body with constant densities 
$\rho_A$ and $\rho_B$ inside and outside the star, the
effective potential has two minima at the field values 
$\phi_A$ and $\phi_B$ satisfying the conditions 
$V_{{\rm eff},\phi}(\phi_A)=0$ and 
$V_{{\rm eff},\phi}(\phi_B)=0$, respectively, 
with mass squared $m_A^2 \equiv V_{{\rm eff},\phi \phi}(\phi_A)$
and $m_B^2 \equiv V_{{\rm eff},\phi \phi}(\phi_B)$.
We define the so-called thin shell parameter \cite{KW}
\begin{equation}
\epsilon_{\rm th} \equiv \frac{\phi_B-\phi_A}{6Q\Phi_c}\,,
\end{equation}
where $\Phi_c$ is the gravitational potential at the surface 
of the body ($\tilde{r}=\tilde{r}_c$). 
If the field $\phi$ is sufficiently heavy such that 
$m_A \tilde{r}_c \gg 1$ and if the body has a thin-shell 
in the region $\tilde{r}_1<\tilde{r}<\tilde{r}_c$
with $\Delta \tilde{r}_c \equiv \tilde{r}_c-\tilde{r}_1 \ll  \tilde{r}_c$, 
the thin-shell parameter is given by 
$\epsilon_{\rm th} \simeq \Delta \tilde{r}_c/\tilde{r}_c
+1/(m_A \tilde{r}_c) \ll 1$ \cite{Tamaki}.
The effective coupling between non-relativistic matter and
the field $\phi$ is $Q_{\rm eff} \simeq 3Q\epsilon_{\rm th}$, 
whose strength can be much smaller than 1 
for $\epsilon_{\rm th} \ll 1$.

The tightest bound on $\epsilon_{\rm th}$ comes from the 
solar system test of the violation of equivalence principle
for the accelerations of Earth and Moon 
toward Sun \cite{KW}.
This constraint is given by \cite{Tsuji08}
\begin{equation}
\epsilon_{\rm th,\oplus}<
8.8 \times 10^{-7}/|Q|\,,
\label{epsi}
\end{equation}
where $\epsilon_{\rm th,\oplus}$ is the thin-shell 
parameter for Earth.
Since the gravitational potential for Earth is 
$\Phi_{c,\oplus}=7.0 \times 10^{-10}$, the condition 
(\ref{epsi}) translates into
\begin{equation}
|\phi_{B,\oplus}|<3.7 \times 10^{-15}\,,
\label{epsi2}
\end{equation}
where we have used $|\phi_{B,\oplus}| \gg |\phi_{A,\oplus}|$.
For cosmologically viable $f(R)$ models, $|\phi_{B,\oplus}|$ is roughly 
the same order as the deviation parameter $m (R_B)$ at the 
Ricci scalar $R_B$ \cite{AT08} (as we will see below).
Hence the condition $m(R_B) \lesssim 10^{-15}$ needs to 
be satisfied in the region around Earth (in which $R_B \gg R_0$).
Cosmologically this means that the parameter $m$ is very much smaller 
than 1 in radiation and deep matter eras.

We consider $f(R)$ models which can be consistent with both cosmological 
and local gravity constraints. They can all be written in the form:
\begin{equation}
f(R) = R - \lambda R_c\,f_1(x)\,, \qquad x\equiv R/R_c\,,
\end{equation}
where $R_c~(>0)$ defines a characteristic value of the Ricci scalar $R$ and 
$\lambda$ is some positive free parameter.

We will study the following models:
\begin{itemize}
\item (A) $f_1(x) = x^p$ \quad ($0<p<1$)\,,
\item (B) $f_1(x) = x^{2n}/(x^{2n} + 1)$ \quad ($n>0$)\,,
\item (C) $f_1(x) = 1 - (1+x^2)^{-n}$ \quad ($n>0$)\,,
\item (D) $f_1(x) = 1 - e^{-x}$\,,
\item (E) $f_1(x) = \tanh (x)$\,.
\end{itemize}
For $\lambda$ of the order of unity, $R_c$ roughly corresponds to the 
scale of the cosmological Ricci scalar $R_0$ today.

The model (A) is characterized by $m=p(r+1)/r$, which 
behaves as $m \simeq p(-r-1)$ during radiation and deep 
matter eras ($r \simeq -1$) \cite{AGPT}.
In the regime $R \gg R_c$ we have $\phi \simeq -(\sqrt{6}/2)m/(1-p)$
with $m \simeq \lambda p(1-p) (R/R_c)^{p-1}$.
In order to satisfy the condition (\ref{epsi2})
the parameter $p$ is constrained to be very small:
$p<3 \times 10^{-10}$ \cite{CT}.

In the regime $R \gg R_c$ the models (B) and (C), proposed 
by Hu and Sawicki \cite{Hu} and Starobinsky \cite{Star} 
respectively, behave as 
\begin{equation}
f(R) \simeq R-\lambda R_c \left[ 1-(R/R_c)^{-2n} \right]\,,
\end{equation}
which corresponds to 
\begin{equation}
m (r)=C(-r-1)^{2n+1}\,,\quad
C=2n(2n+1)/\lambda^{2n}\,,
\label{mrasy}
\end{equation}
with the field value $\phi \simeq -(\sqrt{6}/2)m/(2n+1)$.
Because of the presence of the power $2n+1$ in Eq.~(\ref{mrasy}), 
the condition (\ref{epsi2}) can be satisfied 
even for $n$ and $\lambda$ of the order of unity.
In fact the models (B) and (C) are consistent 
with the constraint (\ref{epsi2}) for $n>0.9$ \cite{CT}.
In these models the deviation parameter $m$ can grow 
to the order of 0.1 today.

The models (D) and (E), proposed by Linder \cite{Linder}
and Tsujikawa \cite{Tsuji} respectively, have 
vanishingly small $m$ in the region $R \gg R_c$, 
but $m$ can grow to ${\cal O}(0.1)$ 
once $R$ decreases to the order of $R_c$
(see Ref.~\cite{Appleby} for a similar model).
In the model (D) one has $m \simeq \lambda (R/R_c) e^{-R/R_c}$ 
for $R \gg R_c$, where $R/R_c \simeq \log (-\sqrt{2/3}\phi/\lambda)$.
The field value $\phi_B$ can be derived
by solving $V_{{\rm eff},\phi} (\phi_B)=0$ 
with $\rho^* \simeq \rho_B$, which gives $R \simeq \rho_B$.
We then find that $\phi_B \simeq -\sqrt{3/2}\lambda e^{-\rho_B/R_c}$.
Since $\lambda R_c$ is of the order of the cosmological density 
$\rho_c^{(0)} \simeq 10^{-29}$\,g/cm$^3$ today,
we have $\phi_B \approx -\lambda e^{-10^5\lambda}$
by using the dark matter/baryonic density $\rho_B \simeq 10^{-24}$\,g/cm$^3$
in our galaxy. As we will see later the existence of a stable de Sitter point 
demands $\lambda>1$, under which the constraint (\ref{epsi2})
is well satisfied. The model (E) also has a similar property.
Thus the models (D) and (E) are consistent with local gravity
constraints for $\lambda$ required for viable cosmology.


\section{Growth indices of matter perturbations}
\label{growthindex}
In this section we study the growth of matter perturbations for 
the viable $f(R)$ models (A)-(E).
We are interested in the wavenumbers $k$ relevant to the 
galaxy power spectrum \cite{Percival}:
\begin{equation}
0.01~h\,{\rm Mpc}^{-1} \lesssim k \lesssim 
0.2~h\,{\rm Mpc}^{-1}\,,
\label{krange}
\end{equation}
where $h=0.72 \pm 0.08$ corresponds to the uncertainty 
of the Hubble parameter today \cite{HST}.
The scales (\ref{krange}) are in the linear regime of perturbations. 
For $k=0.2~h~{\rm Mpc}^{-1}$, non-linear effects are still 
small so that results in the linear regime can be related 
to observations. Non-linear effects increase as we go to smaller scales. 
On the other hand we should remember that observations on the large scale
around $k\sim 0.01~h\,{\rm Mpc}^{-1}$ are not so accurate 
but will be improved in the future. 

We recall that the transition from the ``GR regime'' to the 
``scalar-tensor regime'' occurs at $M^2=k^2/a^2$.
Using Eq~(\ref{mass}) this translates into
\begin{equation}
m \approx (aH/k)^2\,.
\label{mcon}
\end{equation}
This expresses in terms of the quantity $m$ the fact that cosmic scales 
smaller than $\sim M^{-1}$ will be affected by modifications of gravity. 
For viable $f(R)$ models we have presented in the previous section, 
the deviation parameter $m$ increases from the matter era 
to the accelerated epoch.
For larger $k$ (i.e. for smaller scales) the transition occurs 
earlier. The upper bound of the wavenumber in Eq.~(\ref{krange})
corresponds to $k \simeq 600a_0H_0$, where the subscript 
``0'' represents present quantities.
We are interested in the case where the transition to the 
scalar-tensor regime occurred by 
the present epoch (the redshift $z=0$).
This then gives the following condition 
\begin{equation}
m (z=0) \gtrsim 3 \times 10^{-6}\,.
\label{mbound}
\end{equation}
If $m (z=0) \lesssim 3 \times 10^{-6}$ then the linear perturbations 
have been always in the GR regime in the past, so that the models 
are not distinguished from the $\Lambda$CDM model.
We caution that the bound (\ref{mbound}) is relaxed
for non-linear perturbations with $k \gtrsim 0.2\,h$\,Mpc$^{-1}$, 
but the linear analysis  is not valid in such cases.

In our numerical simulations we identify the present epoch 
to be $\Omega_m^{(0)}=0.28$. 

\subsection{Model (A)}
Let us first consider the model (A). In this model the deviation parameter
$m=p(r+1)/r$ corresponds to $m=p/2$ at the de Sitter point ($r=-2$), 
which means that $m (z=0)$ is of the order of $p$.
Hence the condition (\ref{mbound}) for the occurrence of the transition
to the scalar-tensor regime corresponds to 
\begin{equation}
p \gtrsim 6 \times 10^{-6}\,.
\label{pbound}
\end{equation}
%

\begin{figure}
\includegraphics[height=3.0in,width=3.3in]{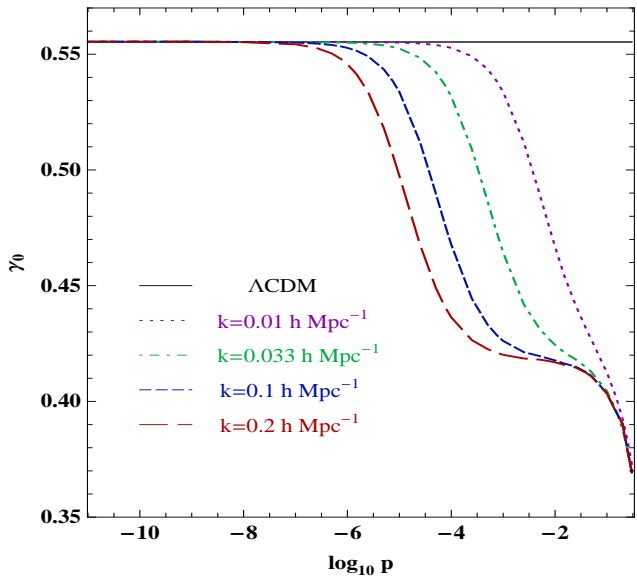}
\caption{\label{figmodela} 
The growth indices $\gamma_0$ today versus the parameter $p$
in the model (A) for four different values of $k$.
Under the local gravity bound $p<3 \times 10^{-10}$, 
the deviation of $\gamma_0$ from that in the $\Lambda$CDM 
model ($\gamma_0 \simeq 0.55$) cannot be seen for 
these wavenumbers.}
\end{figure}

In Fig.~\ref{figmodela} we plot the growth indices $\gamma_0$
of matter perturbations today for four different wavenumbers $k$.
When $p \gtrsim 6 \times 10^{-6}$ the deviation from the value 
$\gamma_0 \simeq 0.55$ of the $\Lambda$CDM model 
can be clearly seen for the modes 
$k \gtrsim 0.2\,h~{\rm Mpc}^{-1}$. 
The dispersion of $\gamma_0$ with respect to the wavenumbers $k$
is especially significant for $10^{-5}<p<10^{-2}$, whereas $\gamma_0$ 
converges to a value around $0.4$
for $p \gtrsim 10^{-1}$.

We recall that local gravity constraints give the bound 
$p<3 \times 10^{-10}$, which is not compatible 
with the condition (\ref{pbound}).
Under this bound the growth indices are very close to the 
$\Lambda$CDM value $\gamma_0 \simeq 0.55$ for 
the wavenumbers (\ref{krange}).
Hence the model (A) cannot be distinguished from the 
$\Lambda$CDM model as long as local gravity 
constraints are respected.

\subsection{Models (B) and (C)}
We shall proceed to the models (B) and (C).
In the region $R \gg R_c$ these models can be described by 
the $m(r)$ curve given in Eq.~(\ref{mrasy}).
In the deep matter era ($r \approx -1$) the deviation parameter $m$
gets smaller for increasing $n$ because of the larger power-law index 
$2n+1$ in Eq.~(\ref{mrasy}). 
For increasing $\lambda$ we also have smaller $m$.

In the model (B) the stability of the late-time de Sitter point 
requires that \cite{Tsuji07}
\begin{equation}
2x_d^{4n}-(2n-1)(2n+4)x_d^{2n}+
(2n-1)(2n-2) \ge 0,
\label{Bmodelcon}
\end{equation}
where $x_d=R_1/R_c$.
 
The parameter 
\begin{equation}
\lambda=\frac{(1+x_d^{2n})^2}{x_d^{2n-1}(2+2x_d^{2n}-2n)}\,,
\label{lamb}
\end{equation}
has a lower bound determined by the condition (\ref{Bmodelcon}).
When $n=1$, for example, one has $x_d \ge \sqrt{3}$ and
$\lambda \ge 8\sqrt{3}/9$.

Similarly the model (C) satisfies 
\begin{equation}
(1+x_d^2)^{n+2}>1+(n+2)x_d^2+(n+1)(2n+1)x_d^4,
\label{Cmodelcon}
\end{equation}
with 
\begin{equation}
\lambda=\frac{x_d(1+x_d^2)^{n+1}}
{2[(1+x_d^2)^{n+1}-1-(n+1)x_d^2]}\,.
\label{Cmodellam}
\end{equation}
When $n=1$ we have $x_d \ge \sqrt{3}$ and 
$\lambda \ge 8\sqrt{3}/9$, which is the same as
in the model (B). For general $n$, however, the 
bounds on $\lambda$ in the model (C) are not 
identical to those in the model (B).
The minimum values of $\lambda$ are of the 
order of unity in both models.

At the de Sitter point the model (\ref{mrasy})
gives $m(r=-2)=C=2n(2n+1)/\lambda^{2n}$, so that 
$m(z=0)$ can be as large as ${\cal O}(1)$
for $n, \lambda$ of the order of unity.
Numerically we find that the deviation parameter $m (z=0)$ 
in the models (B) and (C) is typically smaller than that in the 
model (\ref{mrasy}), but still $m(z=0)$ can be of the order of 0.1.
The deviation parameter $m$ needs to be very much smaller than 1
in the region of high density ($R \gg R_c$) for consistency with 
local gravity constraints.

If the transition characterized by the condition (\ref{mcon}) occurs during the 
deep matter era ($z \gg 1$), one can estimate the critical redshift $z_c$ 
at the transition point. 
We use the asymptotic forms $m \simeq C(-r-1)^{2n+1}$
and $r \simeq -1-\lambda R_c/R$ as well as the approximate relations
$H^2 \simeq H_0^2 \Omega_m^{(0)}(1+z)^3$ and
$R \simeq 3H^2$. The present value of $\rho_{\rm DE}$ may be 
approximated as $\rho_{\rm DE}^{(0)} \approx \lambda R_c/2$.
Hence we have that $\lambda R_c \approx 6H_0^2\Omega_{\rm DE}^{(0)}$, 
where $\Omega_{\rm DE}^{(0)}$ is the DE density parameter 
today. Then the condition (\ref{mcon}) 
translates into the critical redshift
\begin{equation}
z_c=\left[ \left( \frac{k}{a_0H_0} \right)^2
\frac{2n(2n+1)}{\lambda^{2n}} 
\frac{(2\Omega_{\rm DE}^{(0)})^{2n+1}}
{{\Omega_{m}^{(0)}}^{2(n+1)}}
\right]^{\frac{1}{6n+4}}-1.
\label{zc}
\end{equation}
For $n=1$, $\lambda=3$, $k=300a_0H_0$, and 
$\Omega_{m}^{(0)}=0.28$ in the model (C)
the numerical value for the critical redshift is
$z_c=4.5$, which shows good agreement with 
the analytical value estimated by Eq.~(\ref{zc}).
We caution, however, that Eq.~(\ref{zc}) begins to lose 
its accuracy for $z_c$ close to 1.

\begin{figure}
\includegraphics[height=3.0in,width=3.3in]{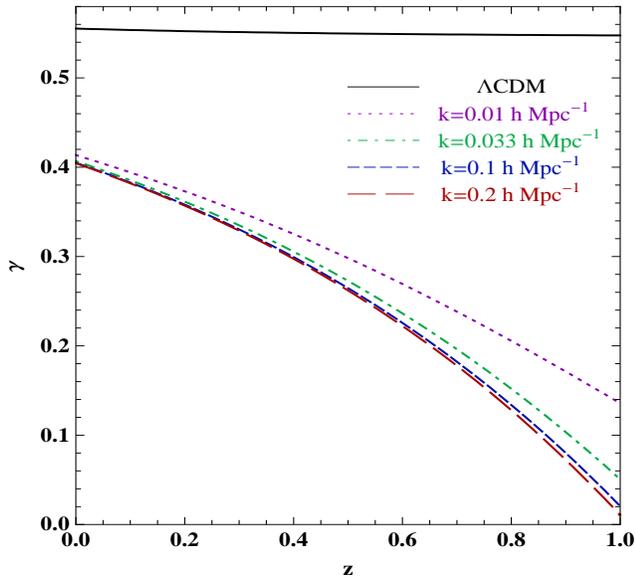}
\caption{\label{figmodelb} 
The evolution of $\gamma$ versus the redshift $z$ in the model (B) 
with $n=1$ and $\lambda=1.55$ for four different values of $k$.
In this case the dispersion of $\gamma$ with respect to $k$
is very small. It is nearly absent for scales 
$k\ge 0.033~h~{\rm Mpc}^{-1}$, so these scales have reached today 
the asymptotic regime $k\gg aM$.
}
\end{figure}

We recall that local gravity constraints give 
the bound $n>0.9$ for both the models (B) and (C).
Meanwhile the conditions (\ref{Bmodelcon})-(\ref{Cmodellam})
provide lower bounds on $\lambda$ for each $n$ 
($\lambda>1.54$ for $n=1$ in both models).
In Fig.~\ref{figmodelb} we plot the evolution of the growth indices 
$\gamma$ in the model (B) with $n=1$ and $\lambda=1.55$ 
for a number of different wavenumbers.
We find a degeneracy of the present value of $\gamma$
around $\gamma_0 \simeq 0.41$ independent 
of the scales of our interest.
In this case the transition redshift corresponds to $z_c=5.2$ and
$z_c=2.7$ for the modes $k=0.1\,h\,{\rm Mpc}^{-1}$ and 
$k=0.01\,h\,{\rm Mpc}^{-1}$, respectively.
At the present epoch these modes are in the ``scalar-tensor'' regime with 
similar growth indices.

\begin{figure}
\includegraphics[height=3.0in,width=3.3in]{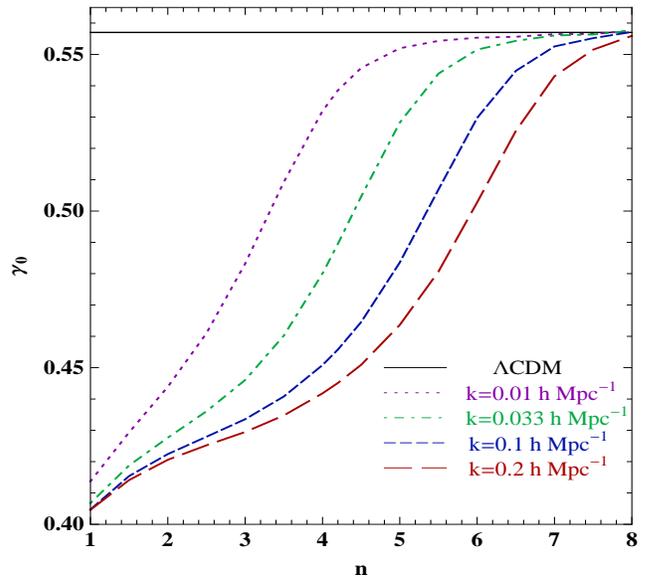}
\caption{\label{figgmodelb1} 
The growth indices $\gamma_0$ today versus $n$
in the model (B) with $\lambda=1.55$ for four different 
values of $k$. The dispersion of $\gamma_0$
occurs in the region $0.42 \lesssim \gamma_0 \lesssim 0.55$
for $2 \lesssim n \lesssim 7$.}
\end{figure}

\begin{figure}
\includegraphics[height=3.0in,width=3.3in]{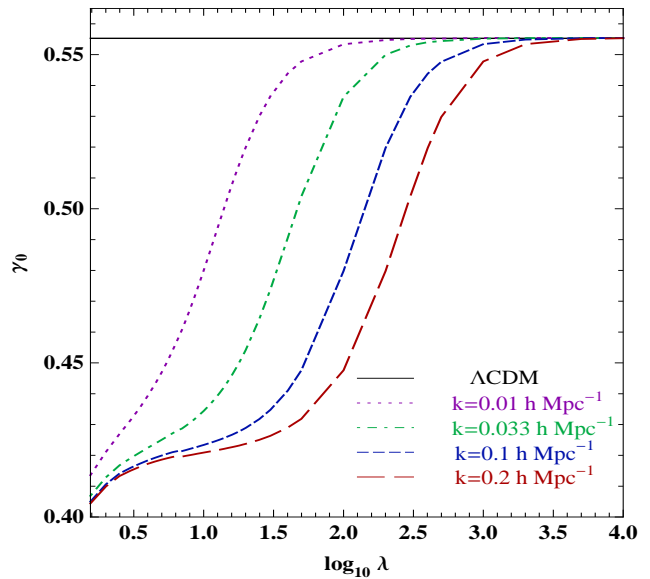}
\caption{\label{figgmodelb2} 
The growth indices $\gamma_0$ today versus $\lambda$
in the model (B) with $n=1$ for four different 
values of $k$. The dispersion of $\gamma_0$ appears 
for $0.5 \lesssim \log_{10} \lambda \lesssim 2.5$.}
\end{figure}

Equation (\ref{zc}) shows that $z_c$ gets smaller for increasing 
$n$. In Fig.~\ref{figgmodelb1} we show the present values 
of $\gamma$ versus $n$ in the model (B) with $\lambda=1.55$ 
for four different wavenumbers.
We find that $\gamma_0$ has a scale dependence in the region 
$0.42 \lesssim \gamma_0 \lesssim 0.55$ for 
$2 \lesssim n \lesssim 7$, while $\gamma_0$
is degenerate around $0.41$ for $n$ close to 1.
This reflects the fact that, for larger $n$, 
the transition redshift $z_c$ gets smaller.
The growth indices are strongly dispersed if the mode $k=0.2\,h\,{\rm Mpc}^{-1}$
crossed the transition point at $z_c>{\cal O}(1)$ and the mode $k=0.01\,h\,{\rm Mpc}^{-1}$
has marginally entered (or has not entered) the scalar-tensor regime by today.
Since $z_c$ decreases for increasing $\lambda$ from Eq.~(\ref{zc}), 
it is expected that the scale dependence of $\gamma_0$ can appear
for larger $\lambda$ than in the case shown in Fig.~\ref{figgmodelb1}
(for fixed $n$). In fact this behavior is clearly seen in 
the numerical simulation of Fig.~\ref{figgmodelb2},
which shows that in the model (B) with $n=1$
the dispersion of $\gamma_0$ occurs 
for $0.5 \lesssim \log_{10}\,\lambda \lesssim 2.5$.
If $\lambda \gtrsim 10^3$, $\gamma_0$ converges to 
the $\Lambda$CDM value $\simeq 0.55$ because 
the modes (\ref{krange}) have not entered 
the scalar-tensor regime by today.

We have also carried out numerical simulations for the model (C) and 
found that the evolution of $\gamma$ is very similar to 
that in the model (B) for the same values of $n$ and $\lambda$.
Let us consider the parameter regions of $(n, \lambda)$ 
for the models (B) and (C) in which the dispersion of $\gamma_0$ occurs
for the wavenumbers (\ref{krange}). 
We can divide the $(n, \lambda)$ plane in three regions:
\begin{itemize}
\item (i) All modes have the values of $\gamma_0$ close to the $\Lambda$CDM value: 
$\gamma_0= 0.55$, i.e. $0.53 \lesssim \gamma_0 \lesssim 0.55$. 
\item (ii) All modes have the values of $\gamma_0$ close to the value
in the range $0.40 \lesssim \gamma_0 \lesssim 0.43$. 
\item (iii) The values of $\gamma_0$ are dispersed in the range $0.40
\lesssim \gamma_0 \lesssim 0.55$.
\end{itemize}

We recall that the first case arises when all scales under consideration 
are close to the asymptotic regime for scales larger today than the 
range of the ``fifth-force''. The second case corresponds to the opposite 
situation. In the third case some of the scales belong to the intermediate 
regime \cite{Gannouji09}. To find out accurately when the asymptotic regimes 
are reached, and what are the values of $\gamma_0$ in the intermediate regime, 
one has to resort to numerical calculations. 

The region (i) is characterized by the opposite of the inequality (\ref{mbound}), i.e.
$m(z=0) \lesssim 3 \times 10^{-6}$. This corresponds to the case in which 
$n$ and $\lambda$ take large values so that $m$ is suppressed.
The border between (i) and (iii) is 
determined by the condition $m(z=0) \approx 3 \times 10^{-6}$.
The region (ii) corresponds to small values of $n$ and $\lambda$, 
as in the numerical simulation of Fig.~\ref{figmodelb}.
In this case the mode $k=0.01\,h\,{\rm Mpc}^{-1}$ at least 
entered the scalar-tensor regime for $z_c>{\cal O}(1)$.

\begin{figure}
\includegraphics[height=3.0in,width=3.3in]{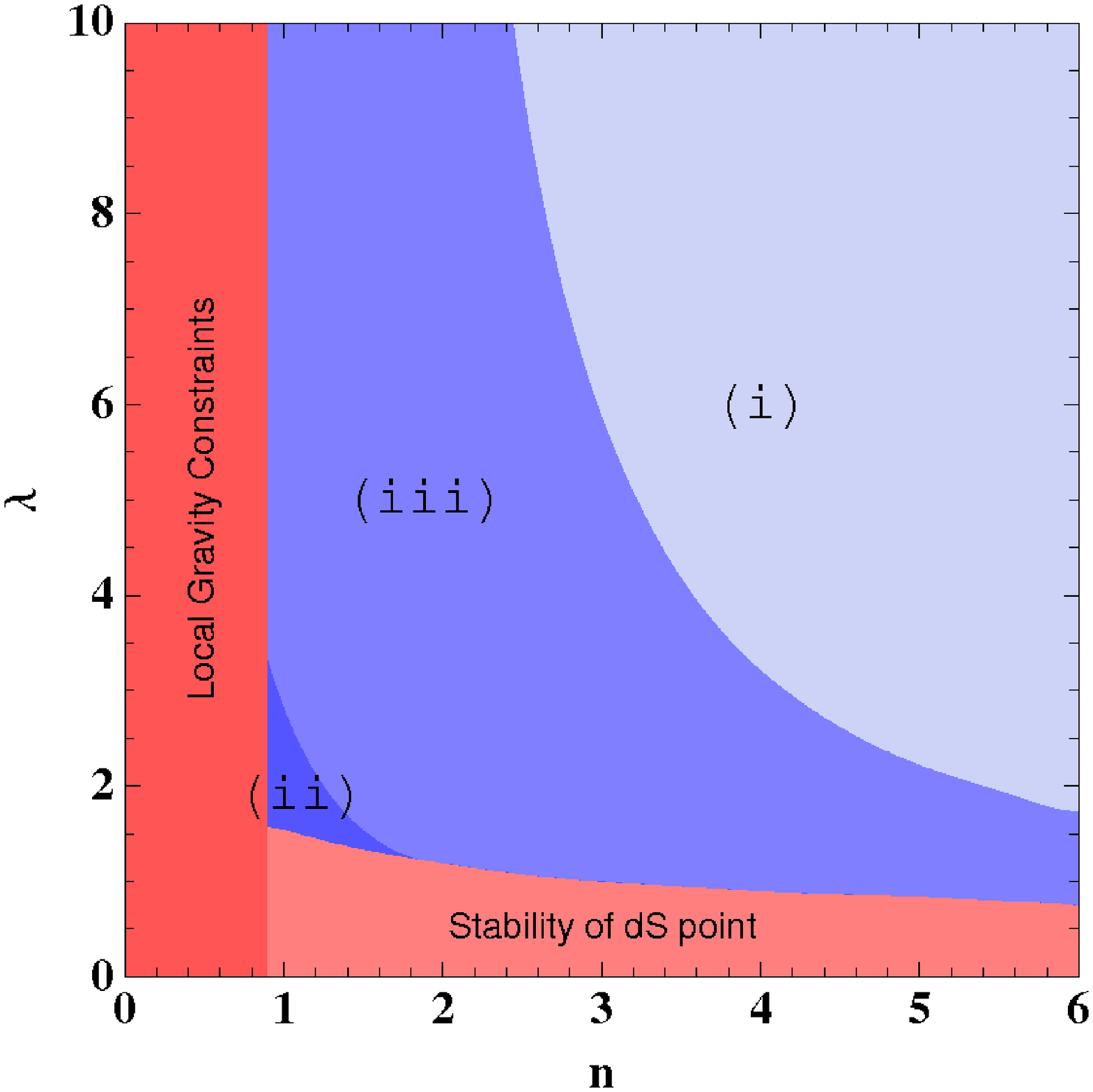}
\includegraphics[height=3.0in,width=3.3in]{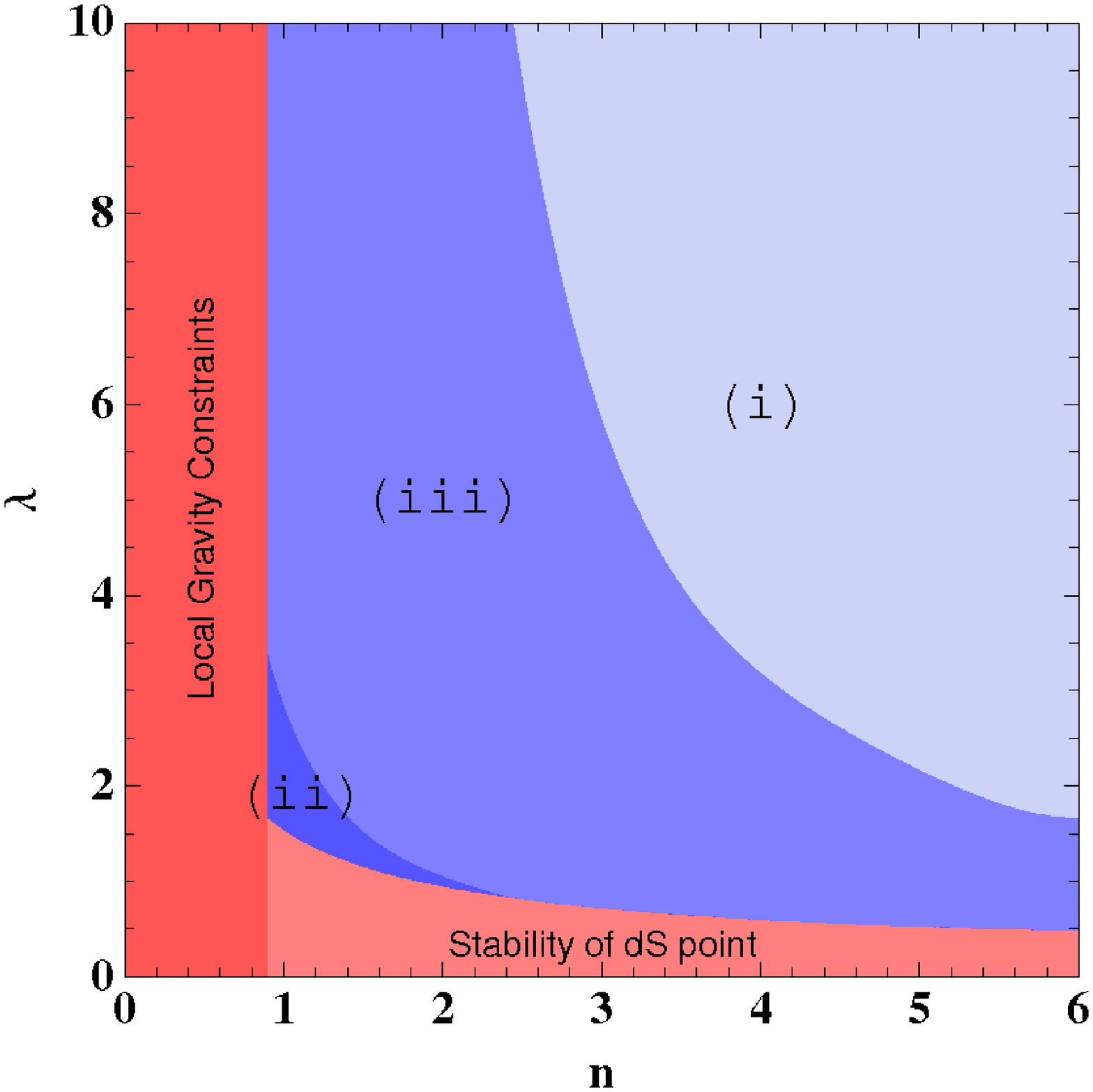}
\caption{\label{ngammaphasek} 
The regions (i), (ii) and (iii) for the model (B) (top)
and for the model (C) (bottom).
The three regions in the model (B) are similar to those in the 
model (C).}
\end{figure}

The regions (i), (ii), (iii) can be found by solving perturbation equations 
numerically. Note that we also have the local gravity constraint $n>0.9$
as well as the conditions (\ref{Bmodelcon}) and (\ref{Cmodelcon}) 
with (\ref{lamb}) and (\ref{Cmodellam})
coming from the stability of the late-time de Sitter point.
In Fig.~\ref{ngammaphasek} we illustrate the regions 
(i), (ii), (iii) for the models (B) and (C), which are quite 
similar in both models.
The parameter space for $n \lesssim 3$ and $\lambda={\cal O}(1)$ 
is dominated by either the region (ii) or the region (iii).
These unusual converged or dispersed spectra can be useful 
to distinguish the $f(R)$ gravity from 
the $\Lambda$CDM model.

\subsection{Models (D) and (E)}
The deviation parameter $m$ in the model (D) is given by 
\begin{equation}
m=\frac{\lambda x e^{-x}}{1-\lambda e^{-x}}\,.
\label{mde}
\end{equation}
In the region $R \gg R_c$ we have that $m \simeq \lambda x e^{-x}$,
which means that $m$ rapidly decreases as we go back to the past.
In the asymptotic past the model (E) has a similar 
dependence $m \simeq 8\lambda x e^{-2x}$.
In both models the parameter $r$ behaves as 
$r \simeq -1-\lambda/x$ for $R \gg R_c$.

For the model (D) the Ricci scalar at the de Sitter point 
($x_d=R_1/R_c$) is determined by $\lambda$, as
\begin{equation}
\lambda=\frac{x_d}{2-(2+x_d)e^{-x_d}}\,.
\label{lammodelD}
\end{equation}
{}From Eqs.~(\ref{mde}) and (\ref{lammodelD}) we find that 
the stability condition $m(R_1)<1$ is satisfied for $x_d>0$.
It then follows from Eq.~(\ref{lammodelD}) that $\lambda$
is bounded to be 
\begin{equation}
\lambda>1\,.
\end{equation}

If the crossing $M^2=k^2/a^2$ occurs during the matter era, 
the transition redshift $z_c$ for the model (D) can be estimated as
\begin{equation}
\frac{2 \Omega_{\rm DE}^{(0)}}{\lambda^2 (1+z_c)^2} 
\exp \left[ \frac{\Omega_{m}^{(0)} \lambda (1+z_c)^3}
{2\Omega_{\rm DE}^{(0)}} \right]=
\left( \frac{k}{a_0H_0} \right)^2\,.
\label{zcest}
\end{equation}
The redshift $z_c$ gets larger for increasing $k$
and for decreasing $\lambda$.
If $k=300a_0H_0$ and $\lambda=1.1$ we have 
$z_c=3.0$ from the estimation (\ref{zcest}).
This is slightly different from the numerical value
$z_c=2.7$ because the transition point is close 
to the onset of the cosmic acceleration.

\begin{figure}
\includegraphics[height=3.0in,width=3.3in]{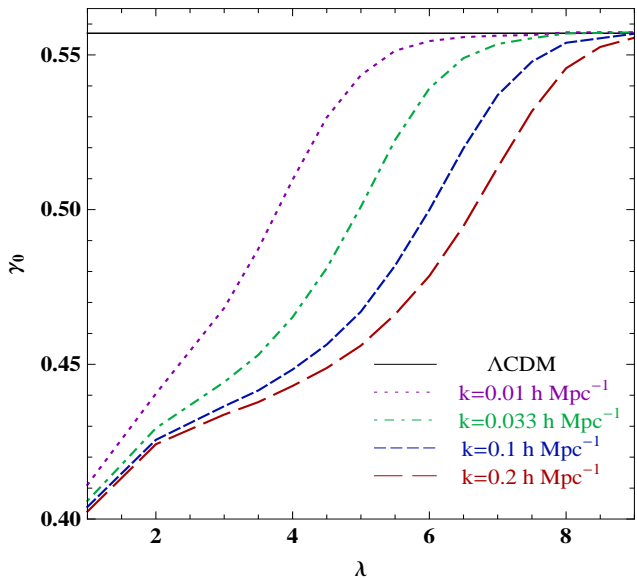}
\caption{\label{figgmodeld} 
The growth indices $\gamma_0$ today versus $\lambda$
in the model (D) for four different values of $k$. We note that 
the transition is much faster in terms of $\lambda$ than in 
model (B) shown in Fig.~\ref{figgmodelb2}.}
\end{figure}

In Fig.~\ref{figgmodeld} we plot the growth indices 
$\gamma_0$ today versus $\lambda$ for four 
different wavenumbers. 
If $\lambda$ is close to 1 then $0.40<\gamma_0<0.42$, 
so that the dispersion of $\gamma_0$ is weak.
The dispersion begins to appear for $\lambda>2$. 
This is associated with the fact that 
the transition redshift gets smaller for increasing $\lambda$.
If the condition $m(z=0) \lesssim 3 \times 10^{-6}$ is satisfied, 
the transition does not occur by today so that 
$\gamma_0$ is close to the $\Lambda$CDM value $0.55$
for the modes (\ref{krange}).
Numerically the present value of $x$ is found to be 
$x_0 \approx 2.2 \lambda$.
Plugging this into Eq.~(\ref{mde}), we find that the condition 
$m(z=0) \lesssim 3 \times 10^{-6}$ translates into $\lambda \gtrsim 8$.
This shows that the dispersion of $\gamma_0$ in the range 
$0.40 \lesssim \gamma_0 \lesssim 0.55$
occurs for $2 \lesssim \lambda \lesssim 8$.
This can be confirmed in the numerical simulation 
of Fig.~\ref{figgmodeld}.
For $\lambda \gtrsim 8$, $\gamma_0$ converges 
to the $\Lambda$CDM value $\simeq 0.55$.

\begin{figure}
\includegraphics[height=3.0in,width=3.3in]{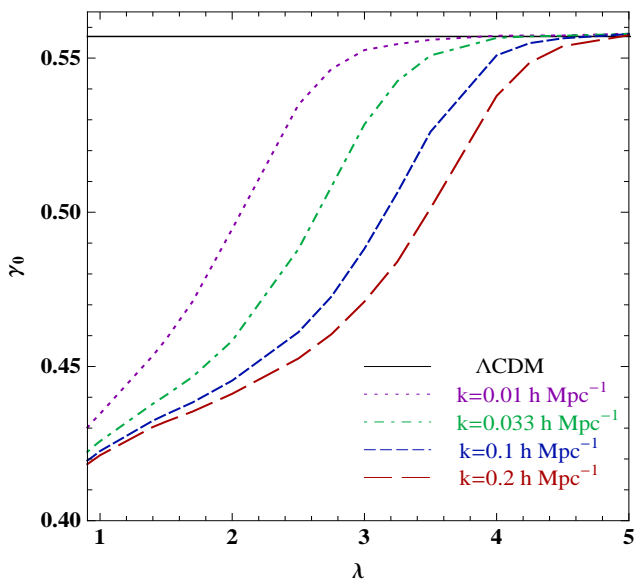}
\caption{\label{figgmodele} 
The growth indices $\gamma_0$ today versus $\lambda$
in the model (E) for four different values of $k$. It is seen 
that the transition for this model is slightly more rapid in terms of 
$\lambda$ than in model (D) shown in Fig.~\ref{figgmodeld}.}
\end{figure}

For the model (E) we have 
\begin{equation}
m=\frac{2\lambda x \tanh (x) [ 1-\tanh^2(x)]}
{1-\lambda [1-\tanh^2(x)]}\,.
\label{zcest2}
\end{equation}
The de Sitter point is determined by the relation
\begin{equation}
\lambda=\frac{x_d \cosh^2 (x_d)}
{2\sinh (x_d) \cosh (x_d)-x_d}\,.
\end{equation}
{}From the stability of the de Sitter point we require 
that \cite{Tsuji07} 
\begin{equation}
\lambda>0.905\,,\quad x_d>0.920\,.
\end{equation}
As in the model (D) the numerical value of $x_0$ 
is about $x_0 \approx 2.2 \lambda$.
It then follows from Eq.~(\ref{zcest2}) that 
the condition $m(z=0) \lesssim 3 \times 10^{-6}$
corresponds to $\lambda \gtrsim 4$,
in which case $\gamma_0$ is degenerate to 
$\gamma_0 \simeq 0.55$ for the modes (\ref{krange}).
In Fig.~\ref{figgmodele} the dispersion of $\gamma_0$ 
can be seen for $\lambda \lesssim 4$.
When $\lambda$ is close to the minimum 
value $0.905$, the growth indices are almost degenerate in the range
$0.42 \lesssim \gamma_0 \lesssim 0.43$.


\section{Conclusions}
\label{conclusions}

In this paper we have studied the dispersion of the growth index 
$\gamma$ of matter perturbations in $f(R)$ gravity models.
We focused on a number of viable $f(R)$ dark energy models 
proposed in the literature that can satisfy cosmological and local 
gravity constraints.
While these models are close to the $\Lambda$CDM
model in the asymptotic past, a deviation from 
the $\Lambda$CDM model appears at late times.
A useful quantity that characterizes this deviation is 
given by $m=Rf_{,RR}/f_{,R}$.
This quantity needs to be very much smaller than 1 
during the deep matter era for consistency with 
local gravity constraints, but a growth of $m$ 
to the present value of the order of 0.1 can be allowed
depending on the $f(R)$ models.

The transition of matter perturbations from the GR regime 
to the scalar-tensor regime occurs at the epoch characterized 
by the condition $m \approx (aH/k)^2$. 
For the wavenumbers $k$ relevant for the observable range of 
the linear matter power spectrum 
($0.01\,h\,{\rm Mpc}^{-1} \lesssim k \lesssim
0.2\,h\,{\rm Mpc}^{-1}$), we require that 
$m (z=0) \gtrsim 3 \times 10^{-6}$ for the occurrence 
of such a transition by today.
For the model (A) this requirement is not compatible 
with local gravity constraints and hence this model 
cannot be distinguished from the $\Lambda$CDM model.

The models (B) and (C) allow for a rapid growth of $m$
from the region $R \gg H_0^2$ [with $m(R) \lesssim 10^{-15}$]
to the region $R \simeq H_0^2$ [with $m(R)={\cal O}(0.1)$].
When $n<3$ and $\lambda={\cal O}(1)$ we find 
two distinct regions widely spread in the $(n, \lambda)$ plane:
the region (ii) in which the present growth indices $\gamma_0$
almost converge to the values around 
$0.40 \lesssim \gamma_0 \lesssim 0.43$ and 
the region (iii) in which $\gamma_0$ are dispersed
around $0.40 \lesssim \gamma_0 \lesssim 0.55$. In the first 
region there is essentially no spatial dispersion of $\gamma_0$, 
in contrast to the second region. These results are summarized 
in Fig.~\ref{ngammaphasek} for the models (B) and (C).

The models (D) and (E) give rise to an even faster evolution 
of $m$ compared to the models (B) and (C). 
The evolution of $\gamma$ depends on the single parameter 
$\lambda$. For the models (D) and (E) the dispersion of $\gamma_0$ 
in the region $0.40 \lesssim \gamma_0 \lesssim 0.55$ is found  
for $2 \lesssim \lambda \lesssim 8$ and 
$1.5 \lesssim \lambda \lesssim 4$, respectively.
The growth indices converge to values around 
$0.40 \lesssim \gamma_0 \lesssim 0.43$ 
for $1<\lambda \lesssim 2$ [model (D)] and 
for $0.905<\lambda \lesssim 1.5$ [model (E)].

We have thus shown that the dispersed or converged 
growth indices with $\gamma_0$ smaller than 0.55
are present in viable $f(R)$ models with 
$m(z=0) \gtrsim 3 \times 10^{-6}$.
If future observations detect such unusually small values 
of $\gamma_0$, this can be a smoking gun for $f(R)$ models.
The presence of some dispersion in future observations could 
be an additional evidence for some of our $f(R)$ models.
We also note that our analysis can be extended to 
scalar-tensor models with couplings $Q$ of the order of 1
between dark energy and non-relativistic matter 
in the Einstein frame \cite{Tsuji08,Gannouji09}.
It will be of interest to investigate the growth of 
matter perturbations and the resulting dispersion of 
$\gamma$ in such theories.

\section*{ACKNOWLEDGEMENTS}
ST thanks financial support for JSPS (No.\,30318802).
DP thanks for hospitality Tokyo University of Science 
where the present project was initiated. 

\end{document}